\documentclass{sig-alternate}
\usepackage{graphicx}
\usepackage{subfig}

\usepackage{xcolor}

\begin{document}
\begin{sloppypar}
\setcopyright{rightsretained}




%

\conferenceinfo{Neu-IR '16 SIGIR Workshop on Neural Information Retrieval}{\\ July 21, 2016, Pisa, Italy}

\title{Toward a Deep Neural Approach for Knowledge-Based IR}
%

\numberofauthors{4} 
\author{
\alignauthor Gia-Hung Nguyen\\
       \affaddr{IRIT, Universit\'e de Toulouse }\\
       \affaddr{UPS 118 Route Narbonne, Toulouse, France}\\
       \email{gia-hung.nguyen@irit.fr}
\alignauthor Lynda Tamine\\
       \affaddr{IRIT, Universit\'e de Toulouse }\\
       \affaddr{UPS 118 Route Narbonne, Toulouse, France}\\
       \email{tamine@irit.fr}
\alignauthor Laure Soulier\\
       \affaddr{Sorbonne Universit\'es, UPMC Univ Paris 06 CNRS, LIP6 UMR 7606, 4 place Jussieu 75005 Paris, France} \\
       \email{laure.soulier@lip6.fr}
\and
\alignauthor Nathalie Bricon-Souf\\
       \affaddr{IRIT, Universit\'e de Toulouse }\\
       \affaddr{Castres, France}\\
       \email{nathalie.souf@irit.fr}
}

\maketitle
\begin{abstract}
This paper tackles the problem of the semantic gap between a document and a query within an ad-hoc information retrieval task. In this context, knowledge bases (KBs) have already been acknowledged as valuable means since they allow the representation of explicit relations between entities. However, they do not necessarily represent implicit relations that could be hidden in a corpora. This latter issue is tackled by  recent works dealing with deep representation learning of texts. With this in mind, we argue that embedding KBs within  deep neural architectures supporting document-query matching would give rise to fine-grained latent representations of both words and their semantic relations.\\
In this paper, we  review the  main approaches of neural-based document ranking as well as  those approaches for latent representation of entities and relations via  KBs. We then propose some avenues to incorporate KBs in deep neural approaches for document ranking. More particularly, this paper advocates that KBs can be used either to support enhanced latent representations of queries and documents based on both distributional and relational semantics or to serve as a semantic translator between their latent distributional representations.
\end{abstract}


\section{Introduction}
Knowledge resources such as ontologies and Knowledge bases (KBs) provide data which are critical to associate words with their senses. Even if word senses cannot be easily discretized to a finite set of entries, numerous works have shown that such resources can successfully bridge the semantic gap between the document and the query within an information retrieval (IR) task \cite{Cao:2005}. More specifically, in this line of work, those resources allow to enrich word-based representations by mapping words to concepts or entities and exploiting symbolic formalized semantic relations (e.g., ``is-a" or ``part-of") between words.  Another way to deal with word senses is to learn from corpora their representations based on the premise of distributional semantics  \cite{mikolov:w2v,Pennington:2014}, also called word embeddings. Numerous recent works in this other line of works learn deep word representations by exploiting the context window surrounding the word. Furthermore, based on the general approach of latent representations of texts, several works attempt to model the relevance scoring of latent representations using deep neural architectures \cite{huang:dssm,severyn:shorttextrank}.\\

In this paper, we argue that combining (1) distributional semantics  learned through deep architectures from the text corpora, and (2) symbolic semantics held by extracted concepts or entities from texts based on digital knowledge, would enhance the learning algorithm of latent representations of queries and documents with respect to the IR task. Thus, we propose two general deep architectures that incorporate a knowledge-based source of evidence in the input layer.   The aim of the first approach is to combine word-based semantics and relational-based semantics in the query/document representations. The learning model attempts to map a term-concept-relation vector of  the query/document to an abstracted representation. Unlikely, the main objective in the second approach is to jointly learn latent representations of the document and the query as well as a semantic translator between these latent entities. Therefore, the objective of the learning algorithm is to map a term-based representation of the query/document and joint concept-relation representation to a low-dimensional semantic representation. \\

The rest of this paper is organized as follows. Section 2 discusses previous works related to neural approaches of ad-hoc IR and for latent representation of KB entities and relations or using KBs for improving latent text representations.  Section 3 presents our approaches for using KBs as part of a deep neural architecture for performing ad-hoc IR. Section 4 concludes the paper and outlines relevant future work in the line of the proposed approaches. \\

                            
\section{Related work}
Deep learning techniques have shown strong performance in many natural language processing and IR tasks. According to the motivation of this paper, we review here how deep neural networks have been leveraged for both document-query matching tasks as well as the representation of KBs.
\subsection{On using Deep Neural Networks in IR}
Recently, many works have shown that deep learning approaches are  highly efficient in several IR tasks (e.g., text matching \cite{bengio:neuralLM,huang:dssm}, query reformulation \cite{Mitra:2015}, or  question-answering \cite{Bordes:2014}). 
More close to our work, we consider in this paper the specific task of text matching and the use of deep neural networks for document ranking. Indeed, deep architectures have been highlighted as  effective in the  discovery of hidden structures underlying plain text modeled through latent semantic  features.  
\begin{figure}
\centering
\includegraphics[width=0.4\textwidth]{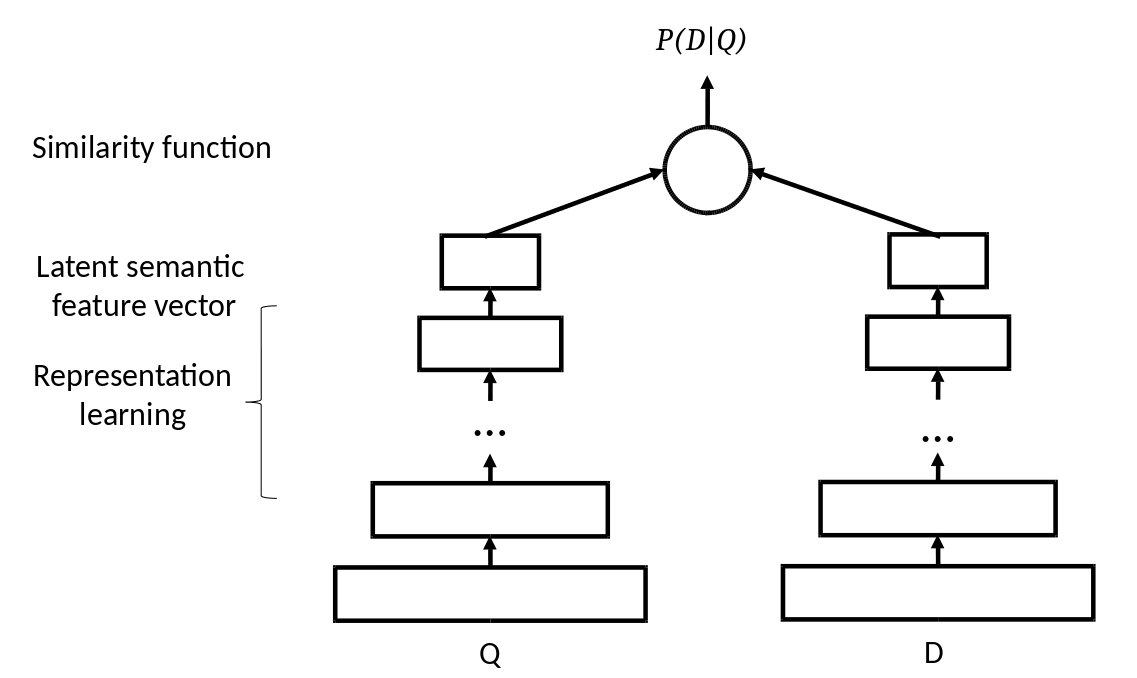}
\caption{General architecture of the DSSM network.}
\label{DSSM}
\end{figure}

We distinguish between two types of neural IR models according to the learning and leveraging approaches of  distributed representations of text. The first category of work uses distributed representations to exploit text dependence within a well-known IR model, such as language models \cite{bengio:neuralLM,Mikolov:documents}.
Also, Mitra  \cite{mitra:dual} has recently proposed a model that leverages the dual word embeddings to better measure the document-query relevance. By keeping both input and output projections of \texttt{word2vec} \cite{mikolov:w2v}, this \textit{Dual Embedding Space Model} allows to leverage both the embedding spaces to acquire richer distributional relationships. The author has demonstrated that this model is able to better gauge  the document \textit{aboutness} with respect to the query.

The second category of works, which knows a keen interest in the recent years, consists in end-to-end scoring models that learn the relevance of document-query pairs via  latent semantic features \cite{huang:dssm,shen:clsm} by taking into consideration the retrieval task objective. These models, also called Deep Semantic Structured Model (DSSM), have been introduced  by Huang et al. \cite{huang:dssm} and are reported to be strong  ones in web search task.
In this approach, the query and the document are first modeled as two high dimensional term vectors (e.g., bag-of-words representation). Through a feed-forward neural network, as shown in Figure \ref{DSSM},   the DSSM learns a representation of these entities (namely document and query) so as to obtain a low-dimensional vector projected within a latent semantic space. Then, the document ranking is trained, always within the DSSM architecture,  by the maximization of  the conditional likelihood of the query given the document.
More particularly, the authors estimate this conditional likelihood by a softmax function applied on the cosine similarity between the corresponding semantic vector of documents and queries.
Moreover, to tackle the issue of large vocabularies surrounding long texts and to enable large-scale training, the authors have proposed the word hashing method which transforms the high-dimensional term vector of the query/document to a low-dimensional letter-trigram vector. This lower dimensional vector is then considered as input of the feed-forward neural network. 

As an extension of the DSSM proposed in \cite{huang:dssm}, Shen et al. \cite{shen:clsm} propose to consider word-trigram vectors enhanced by a word hashing layer (instead of word hashing on the basis of bag-of-words) to capture the fine-grained contextual structures in the query/document. Accordingly, the end-to-end scoring model is impacted, leading to a convolutional-pooling structure, called Convolutional Latent Semantic Model (CLSM). 

In the same mind, Severyn and Moschitti \cite{severyn:shorttextrank} present another convolutional  neural network architecture to learn the optimal representation of short text pairs as well as the similarity function. Given a pair of sentences modeled as a matrix of pre-trained word embeddings, this model first learns their intermediate feature representation by applying convolution-pooling layers on each sentence. A similarity score of this  intermediate representation of the document and the query is computed and enhanced then by additional features (e.g., query-document word/IDF overlap). This richer representation is plugged into a fully connected layer that classifies whether or not the document is similar to the query. Another convolutional architecture model for matching two sentences is proposed in \cite{hu2014convolutional}.
Instead of relying on theirs semantic vectors, the authors use a deep architecture with multiple convolutional layers to model an interaction between plain texts (i.e. the co-occurence pattern of words across two texts). The proposed model allows to represent the hierarchical structures of sentences and to capture the rich matching patterns at different levels.

\subsection{Leveraging knowledge graph for distributed representations}
The potential of semantic representations of words learned through a neural approach has been introduced in \cite{mikolov:w2v,Pennington:2014}, opening several perspectives in natural language processing and IR tasks. Beyond words, several works focused on the representation of sentences \cite{Mikolov:phrases}, documents \cite{Mikolov:documents}, and also knowledge bases (KBs) \cite{Bordes:2013,xu:rc-net}. Within the latter work focusing on KBs, the goal is to exploit concepts and their relationships to obtain a latent representation of the KB.  While some work focused on the representation of relations on the basis of triplets belonging to the KB \cite{Bordes:2013}, other work proposed to enhance the distributed representation of words for representing their underlying concepts by taking into consideration the structure of the KB graph (e.g., concepts in the same category or their relationships with other concepts) \cite{faruqui:retrofitting,xu:rc-net,yu:RCM}.

A first  work \cite{faruqui:retrofitting} proposes a ``retrofitting" technique consisting in a leveraging of  lexicon-derived relational information, namely adjacent words of concepts, to refine their associated word embeddings. 
 The underlying intuition is that adjacent concepts in the KB should have similar embeddings while maintaining most of the semantic information in their pre-learned distributed word representations.
For each word, the retrofitting approach learns its new representation by minimizing both (1) its distance with the representation of all connected words in the semantic graph and (2) its distance with the pre-learned word embedding, namely its initial distributed representation. 

\begin{figure*}[t]
\center
\hspace{-2cm}
     \begin{minipage}[c]{.50\linewidth}
\subfloat[Enhanced representation using KB for IR \label{enhanced}]{
    \includegraphics[scale=0.18]{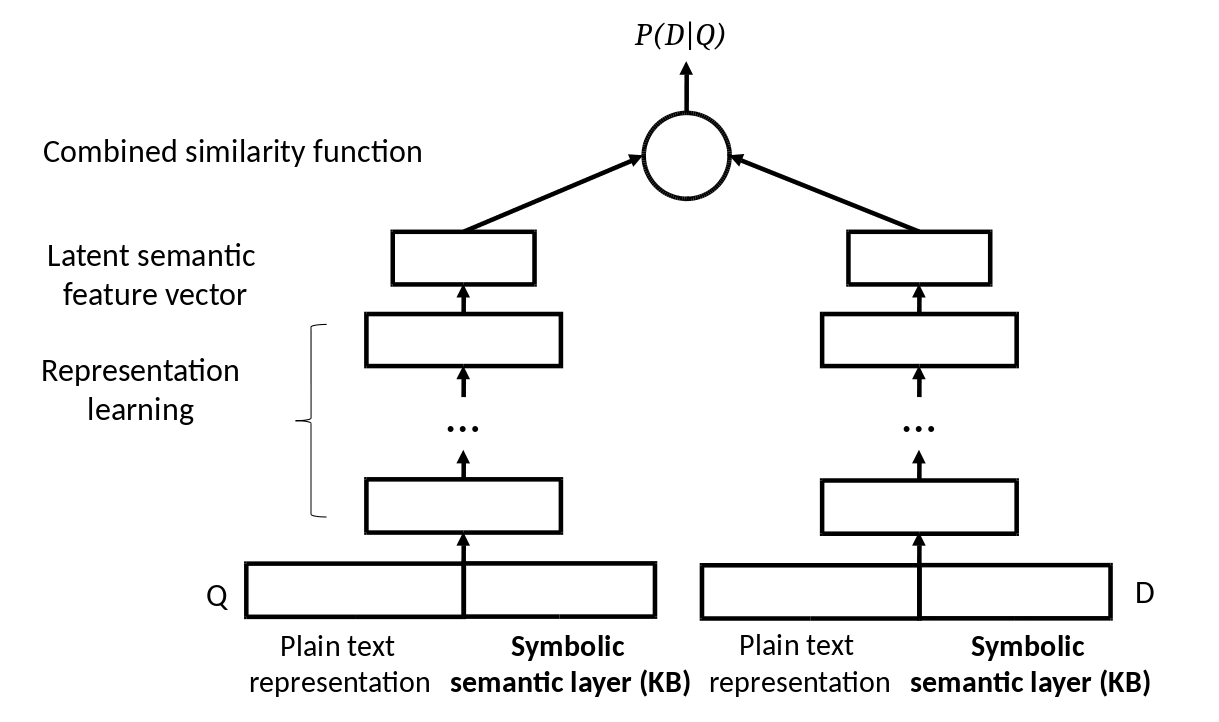}
}
     \end{minipage}  \hspace{-0.5cm}
 \begin{minipage}[c]{.4\linewidth}
\subfloat[The KB translation model \label{translation}]{
    \includegraphics[scale=0.18]{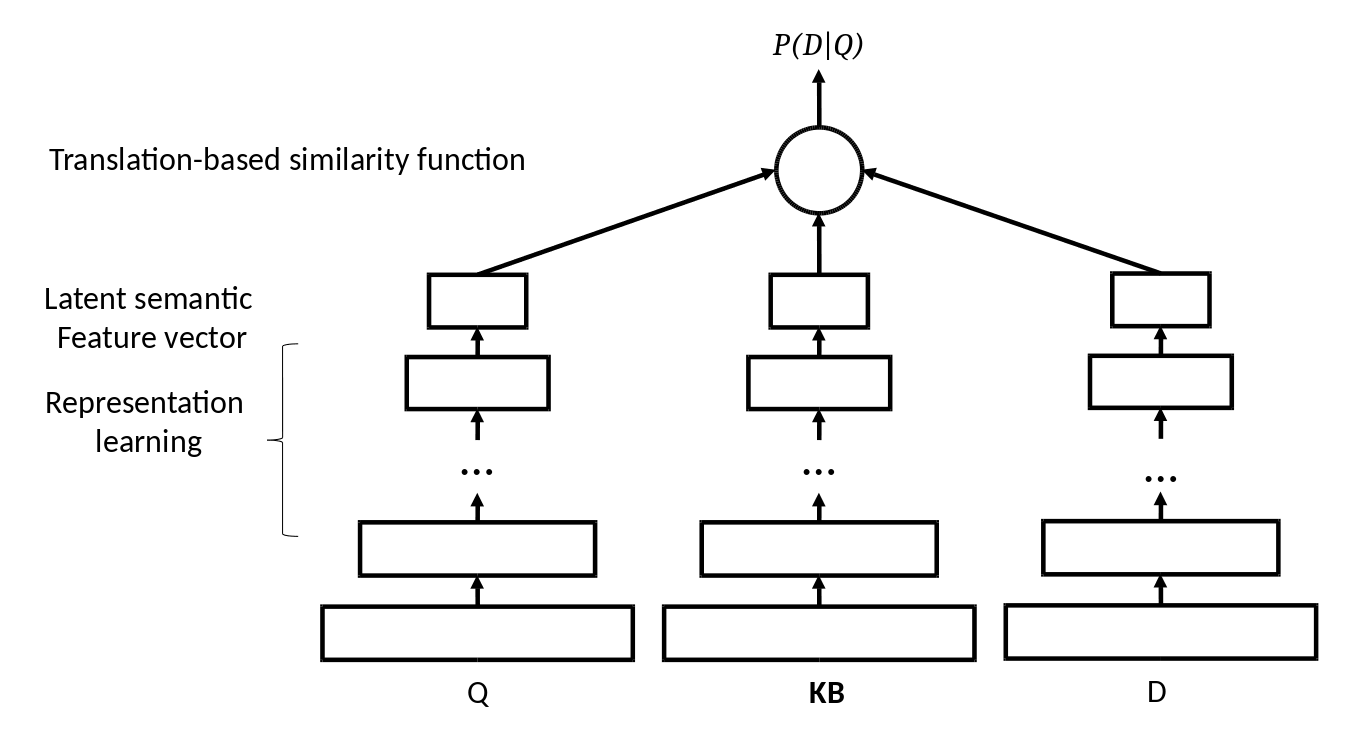}
}
    \end{minipage}   
     \vspace{-0.3cm}
      \caption{Overview of approaches aiming at leveraging KB in DSSM architectures} 
      \vspace{-0.2cm}
      \label{approaches}
\end{figure*}

In contrast to \cite{faruqui:retrofitting}, other work \cite{xu:rc-net,yu:RCM} proposes an end-to-end oriented approach that rather adjusts the objective  function of the neural language model. 
For instance, Xu et al.  \cite{xu:rc-net} propose the RC-NET model that leverages the relational and categorical knowledge to learn a higher quality word embeddings. This model  extends the objective function of the skip-gram model  \cite{mikolov:w2v}  with two regularization functions based on relational and categorical knowledge from the external resource, respectively. While the relational-based regularization function characterizes the word relationships which are  interpreted as translations in latent semantic space of word embeddings, the categorical-based one  aims at minimizing the weighted distance between words with same attributes. With experiments on text mining and NLP tasks, the authors have reported that combining these two regularization functions allows to significantly  improve the quality of word representations.
In the same mind, Yu et al.  \cite{yu:RCM} propose a relation constrained model (RCM) that extends the CBOW model \cite{mikolov:w2v}  with a function based on prior relational knowledge issued from an external resource. 
Thus, the final objective of the model is to learn the pure distributed representation in the text corpus and also to capture the semantic relationship between words from external resources.

In addition to word similarity tasks, the literature review shows that KBs are also exploited in question-answering tasks. For instance,  Bordes et al. \cite{Bordes:2014} exploit a KB to learn the latent representations of  questions and candidate answers. The latter is  modeled as a subgraph built by a sophisticated inference procedure that captures the relationship of the question object with the candidate answer as well as  its related entities.

We have described in this section two branches of work. The first one investigates the use of a deep neural network within a document-ranking matching process, often performed without external features, while the second one  exploits KBs to learn a better distributed representation of words or concepts. In the next section, we will show how KBs could be leveraged within the deep neural network architecture for the document-ranking task.


\section{Toward leveraging KB for neural ad-hoc IR}
The reported literature review clearly highlights the potential of neural networks in one hand and the benefit of KBs, in the other hand, for ad-hoc search tasks. We believe that the integration of an external resource within a document-query neural matching process would allow benefiting from the symbolic semantics surrounding concepts and their relationships. Accordingly, such approach would impact the representation learning that could be performed at different levels. As illustrated in Figure \ref{approaches}, we suggest using a deep neural approach to achieve two levels of representations: 1) an enhanced knowledge-based representation of the document and the query and 2) a distinct representation of  the document and the query surrounding by a third KB-based representation aiming at improving the semantic closeness of document and query representations.\\
While in the first approach, a KB is used as a mean of document and query representation enhancement, the KB is exploited in the latter approach as a mean for document-query translation.

\subsection{Leveraging enhanced representations of text using KB for IR}
The first approach that we suggest for integrating KB within a deep neural network focuses  on an enhanced representation of documents and queries as illustrated in Figure~\ref{enhanced}. While a naive approach would be to exploit  the concept embeddings learned from the KB distributed representation \cite{faruqui:retrofitting,xu:rc-net}  as input of the deep neural network, we  believe that a hybrid representation of the distributional semantic (namely, word embeddings)  and the symbolic semantics (namely, concept embeddings taking into account the graph structure) would allow enhancing the document-query matching. Indeed, simply considering concepts belonging to the KB may lead to a  partial mismatch with the text of queries and/or documents \cite{Cao:2005}. 

With this in mind, the document and query  representations  could be enhanced with a symbolic semantic layer expressing the projection of the plain text on the KB with the consideration of concepts and their relationships within the KB. On one hand, the  representation of the plain text might be, as used in several previous work, a high-dimensional  vector of terms \cite{huang:dssm,shen:clsm} or of their corresponding word embeddings  \cite{severyn:shorttextrank}. 
On the other hand, the semantic layer could be built by the representation of  concepts (and their relationships) extracted from the  plain text  through a concept embedding \cite{faruqui:retrofitting} or a richer embedding representation of a KB sub-graph, as suggested in \cite{Bordes:2014}. The latter presents the advantage to model the compositionality of concepts within the document. 
Similarly to previous approaches  \cite{huang:dssm,severyn:shorttextrank,shen:clsm}, the enhanced representations of both document and query would be transformed into low-dimensional semantic feature vectors used within a similarity function.

\subsection{Using KB translation model for IR}

While the first model exploits knowledge bases to enhance the representation of a document-query pair and their similarity score, an alternative approach consists in a ranking model based on the translation role of the knowledge resource.  As illustrated in Figure \ref{translation}, this second approach aims to take external knowledge resources as a third component of the deep neural network architecture. Intuitively, this third branch could be considered as a pivotal component bridging the semantic gap between the document and the query vocabulary. Indeed, the knowledge resource is here seen as a mediate component that helps to translate the deep representation of the query towards the deep representation of the document with respect to the ad-hoc IR task. 

More practically, the model would consider  three initial entities (namely the document, the query, and the knowledge resource) as inputs. Whether modeled as plain text vector or word embedding matrices, the translation input should be an extraction from the KB characterizing the semantic relationship between the document and the query through their symbolic semantics in the KB (e.g., the embedding of concepts extracted in common in both entities). Then, with a deep architecture, the model will learn the raw representation as a latent semantic feature vector for each entity (document, query, and knowledge-based bridge). Note that in this approach, the  representations of a document-query pair and the representation of the knowledge-based translation vector are learned in the same continuous embedding space.
Then, with the intuition that the KB plays the role of a mediation component, the model will learn the similarity of a document-query pair with a scoring function that takes into account the translation role of the knowledge-based bridge (e.g., vector or matrix translation as done in \cite{Clinchant:2006}).

                
\section{Conclusions}
In this paper, we addressed the emergence of deep learning 
in ad-hoc IR tasks as well as the representation learning approach  of words surrounded by external KB. Following previous work in IR highlighting the benefit of the consideration of the semantic in IR, we have suggested two approaches that leverage  external semantic  resources to improve a text retrieval task within deep structure neural networks.  More particularly, we explained how  KB could be integrated within the representation learning, either through  an  enhanced knowledge-based representation of the document and the query or as a translation representation bridging the semantic gap between the document and the query vocabulary.
We outline that we particularly focused on the DSSM architecture but that our positions could fit with other deep neural network architectures, e.g. recurrent or memory networks \cite{Palangi:2015}. \\ 
We hope that this proposal would support researchers in their future work related to ad-hoc IR as well as other search tasks such as question-answering or entity retrieval. All of these tasks would benefit from combining both distributional and knowledge-based latent representations of texts within the relevance scoring process. 


\bibliographystyle{abbrv}
\bibliography{NeuIR}  
\end{sloppypar}
\end{document}